\documentclass[12pt,halfline,a4paper]{article}

\usepackage{natbib}
\newcommand{\eg}{e.\,g., }
\newcommand{\ie}{i.\,e., }

\newcommand{\ab}[1] {{\color{blue}  ab:  #1}}
\newcommand{\sa}[1] {{\color{magenta}  sa:  #1}}

\usepackage{tikz}
\usepackage{url}
\usetikzlibrary{arrows,calc,intersections,decorations.pathmorphing}

\usepackage{amsmath}
\numberwithin{table}{section}
\numberwithin{figure}{section}

\usepackage{cleveref}

\providecommand{\keywords}[1]
{
  \small	
  \textbf{\textit{}} #1
}

\begin{document}


\noindent

\noindent

\noindent
        {\Large{Non-Verbal Vocalisations and  their Challenges: Emotion, Privacy, Sparseness, and Real Life }
} \\

\noindent
{\Large Anton Batliner}\\
{\Large Shahin Amiriparian}\\
{\Large Bj\"{o}rn W.\ Schuller}



\begin{quote}
\begin{footnotesize}
	\textit{\textbf{Ah}, Rosencrantz! Good lads, how do ye both?}  Shakespeare, Hamlet \\
	 \textit{\textbf{Ah}! Now I've done Philosophy,...} 
    Goethe, Faust I  \\
    \textit{Okay. Uhm, for me, \textbf{ah}, ooh, I would say ...} 
    Allen, Manhattan \\ ~~\\
  \end{footnotesize}
\end{quote}




\begin{abstract}
   Non-Verbal Vocalisations (NVVs) are short `non-word' utterances  without proper linguistic (semantic) meaning but conveying  connotations -- be this emotions/affects or other paralinguistic information. We start this contribution with a  historic sketch: how they were addressed in psychology and linguistics in the last two centuries, how they were neglected later on, and how they came to the fore with the advent of emotion research. We then give an overview of types of NVVs (formal aspects) and  functions of NVVs, exemplified with the  typical NVV  \textit{ah}. 
   Interesting as they are, NVVs come, however, with a bunch of 
   challenges 
   that should be accounted for: 
   Privacy and general ethical considerations prevent them of being recorded in real-life (private) scenarios to a sufficient extent. Isolated, prompted (acted) exemplars do not necessarily model NVVs in context; yet, this is the preferred strategy  so far when modelling NVVs,  especially in AI. To overcome these problems, we argue in favour of corpus-based approaches. This guarantees a more realistic modelling; however,  we are still faced with privacy and sparse data problems. 
\end{abstract}


\section{Introduction}
\label{sec:introduction}

At first sight, the phenomena constituting the first part of the title  -- Non-Verbal Vocalisations (NVVs) -- are easy to explain and understand, when we aim at a prototypical form, namely  \emph{affect bursts} that are defined by  \cite{Scherer94-AB}   as  ``very brief, discrete, non-verbal expressions of affect in [...] voice as triggered by clearly identifiable events.'' It is a short cry/interjection/yell/exclamation, as an expression of a strong emotion out of a small set of basic emotions such as fear or joy.  Given the whole context -- the triggering event like very sad news or a frightening animal, the concomitant facial expression,  the specific segmental and prosodic form of the vocalisation, and  the linguistic and non-linguistic context before and after, it is easy to interpret:  When seeing the movie ``Psycho'', we cannot but conceive the cry of Janet Leigh in the iconic shower scene when facing Antony Perkins with the knife as expression of mortal fear. Yet, it is getting more difficult when we take away one or more signals: we can mute the film, we can only present the audio stream, or we can only present a very short clip with only video or audio. And we can broaden the view, from  such prototypical affect bursts to all kinds of similar acoustic events that express not only strong emotions but all kinds of more or less pronounced affective states; and to NVVs that by themselves are not affective but  can always trigger affect on part of the interlocutor. Then, neither the acoustic \emph{form} might be unequivocal nor the affective or non-affective \emph{function} might be  easily deciphered.

This intrinsic ambiguity is not special: Not only non-verbal but also verbal expressions (speech) are normally produced in a context that contributes to and alters its semantic and pragmatic meaning. Yet, NVVs are special because the other  problems  addressed in the second part of the title apply for them in a specific way:
They are often taken as indicating \emph{emotions} but most of the time, they may not; and if they do, \emph{privacy} considerations prevent them of being recorded. This means in turn massive data \emph{sparseness}, especially for the emotional NVVs; all this results in more or less artificial data instead of \emph{real life} data to be used in research, leading, in turn,  to questionable validity.




In this contribution, we take an empirical stance towards our topic which can be described as `open microphone scenario' -- or more generally, \emph{open world scenario}: As target, we imagine a machine (it might be called `a machine learning device', or `AI') that records and analyses human `vocal interactions' (\ie speech and NVVs).
This is not confined to but normally means Human-Human-Interaction (HHI) or nowadays as well Human-Computer-Interaction (HCI).  
Our machine has to detect NVVs and to recognise/disambiguate which (communicative) function they have. 
The microphone captures NVVs the same way as words, as an acoustic event on the time axis, between words  or in isolation. 
We will  use the term \emph{corpus} when the data were recorded within such an open world scenario, and  the term  \emph{database} for any other collection of data that are more or less \emph{constructed}, \ie the researcher defines, prompts and/or selects the data, by that creating a \emph{closed world scenario}.






\section{History and Definitions}
\label{sec:history-and-definitions}

\begin{figure*}[tb!]
	\begin{center}
 \vspace{-2cm}
		\includegraphics[width=1.0\textwidth]{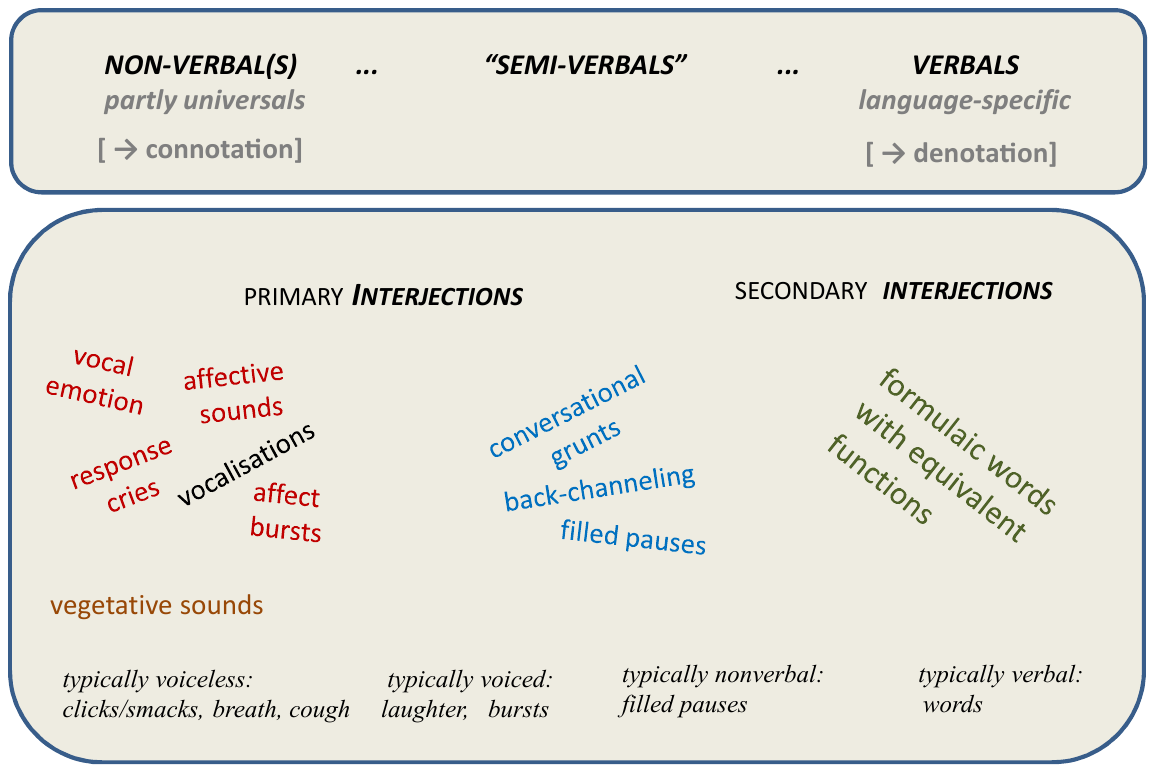}
\vspace{-1cm}
		\caption{
  Types of NVVs
  }
		\label{fig:NVV-types}
	\end{center}
\end{figure*}

 NVVs have always been seen as something special: 
Classic grammarians spoke of \emph{interjections} \citep{Ameka92-ITU}, described by  \cite{Mueller62-LOT} as  \textit{the outskirts of real language}. 
In classic psychology,   \cite{Darwin72-TEO}
 reasoned about the physiological causes for \textit{oh} and \textit{ah} indicating surprise and/or pain; 
 in the same vein, for \cite{James84-WIA}, emotions always have a bodily expression.
Wundt 
used  the term \textbf{primary interjections}, \ie 
voice productions of men and animals when they  precede verbal language or go over into it; they are substituted by \textbf{secondary interjections} dressed up in linguistic form
\citep[p307ff]{Wundt04-V}.
%
%
He further elaborates on \textit{high vocal sounds for aroused affect}, and \textit{lower vocal sounds for depressed feelings}.
In  modern linguistic theories -- past the time of Jespersen and Bloomfield, these `primary interjections' had rather a wallflower existence \citep{Ameka92-ITU}. Normally, they were considered to be peripheral  \citep{Norrick14-I} or not addressed at all \citep{Jensen19-IIS}. 
The same neglect can be seen at the beginning of automatic speech recognition (ASR) that
took them rather as  `garbage',  modelled as a `waste paper basket category'.
%
%
\cite{Schroeder03-ESO,Schroeder03-SAE} rephrases the definition by \cite{Scherer94-AB} given above: Affect bursts are 
\textit{short emotional non-speech expressions conveying a clearly identifiable emotional meaning} comprising both clear non-speech sounds and interjections with a phonemic structure (\textit{wow}) but excluding `verbal' interjections that can occur as a different part of speech (like \textit{Heaven!} or  \textit{No!}). 
This tripartition is given in \Cref{fig:NVV-types}:
\textbf{non-verbals}, \textbf{semi-verbals}, and \textbf{verbals}.  
Clear NVVs are expressions that do not necessarily follow the language-specific phonotactic rules 
that define the permitted combinations of phonemes, 
such as \textit{pfft}; semi-verbals might be conceived as  non-words within the phonotactic constraints such as  \textit{gee} or German \textit{igitt}; formulaic words with both specific meaning and function such as  \textit{my goodness} are verbal. 
Whereas non-verbals which can be sometimes conceived as universals convey  pure connotations, language-specific verbals combine (and normally override) the original denotation with a connotation beyond the literal meaning. 

On the one hand, phylo- and ontogenetically,  NVVs  are primary interjections, \ie before speech and language have been developed, and thus in a way `special'. 
On the other hand, they can be modelled the same way as words because we find them between words
-- albeit sometimes, not corresponding to the native phonotactics, and as `having prosody', the same way as (concatenated) words do have. 
An interlocutor can understand or misunderstand them as affective or not, and misunderstandings can be more or less critical.
NVVs behave like words: Syntagmatically, they are found at specific positions in an utterance; paradigmatically, they can be replaced by words and by that, by secondary interjections. Both primary and secondary interjections  do not need any context, \ie they can stand alone, the same way as a 
single word, an elliptic construction, or a full sentence can stand alone, functioning as a conversational turn.
%

Amongst the more traditional grammarians in the 19th century and the first decades of the last century, see \cite{Sapir21-LAI,Bloomfield33-L}, there only was some general interest in NVVs that turned into neglect in modern linguistic theories and, eventually, narrowed down onto a strong focus on a specific combination of form and function (`affect bursts') on the one hand, in affective science.
On the other hand, in linguistics, pragmatics, and in the field of Non-Verbal Communication, NVVs were dealt with as pragmatic/conversational phenomena.

As often,  terms denoting the same or similar phenomena  overlap somehow and are based on different fields and taxonomies:
\emph{affect bursts} and their synonyms are based on emotion science and delimit the extensional definition. 
`Verbals', `semi-verbal', and `non-verbals' (see \Cref{fig:NVV-types})   are defined linguistically (lexically, morphologically, and phonologically)   and  denote a wide range of functions.
%
%
We can define speech (spoken language) as [+vocal,+verbal] and text (written language) as [-vocal,+verbal];
non-verbal vocalisations are -- as the term indicates -- [+vocal,-verbal]. 
The large field of Non-Verbal Communication \citep{Burgoon22-NC} covers everything else, \ie [-vocal,-verbal], especially body distances and body kinesis (facial expression, head movements, eye behaviour, gestures, postures, gait).
A definition and a separation of the two fields paralinguistics and non-verbal communication, based on these feature values, is given in \cite{Batliner24-P}; see as well \cite{Schuller14-CPE}.

  

The  quotes at the beginning  illustrate the [+vocal,-verbal] phenomena we want to focus on  and their ambiguities; in the [-vocal,+verbal] modality, they are written  as \textbf{\textit{ah}}. Hamlet wants to indicate positive surprise -- on the surface but in fact, he despises the two guys, whereas Faust  wants to indicate frustration or even despair; note that in the German original \textit{Habe nun, \textbf{ach}, Philosophie ...}, substituting \textit{ach} with \textit{ah} would have been possible but less likely. We only can disambiguate the affective meaning of these NVVs by knowing about the (linguistic and pragmatic) context and/or when they are produced with different prosody. When substituting the NVV \textit{ah} with verbals, for Hamlet, we could choose (and by that, disambiguate) \textit{glad to see you} but not for Faust, where we would decide for \textit{alas} or similar expressions. 
In contrast, the \textit{ah} from the third quotation can be substituted by words such as \textit{I mean} but, as a pure hesitation marker, not semantically disambiguated; we know that it is a pure hesitation because of the linguistic context before and after, not because of its phonetic form that could as well indicate slight surprise.

%
%

\section{Types of NVVs}
\label{sec:types}




The term \emph{affect bursts} and its   synonyms given to the left of \Cref{fig:NVV-types} are rooted in psychology and characterise pure connotations:
Wundt's primary interjections named differently such as
\emph{vocal emotion}, 
\emph{affective sounds}, 
\emph{affect bursts} \citep{Scherer94-AB,Schroeder03-ESO}, 
\emph{response cries} \citep{Goffman78-RC},
or 
\emph{vocalisations} \citep{Holz21-TPR,Holz22-TVI}. They need not but can follow the phonotactic rules of a given language: a bilabial trill does not belong to the phomenes of most languages,  but \emph{ah} expressing surprise or disappointment does.
A special category are \emph{vegetative sounds} \citep{Trouvain12-CNV} such as snoring, swallowing, coughing, or yawning. Some of them are purely non-voluntary (for instance sneezing), some of them can be employed as well voluntarily, with some connotative meaning, \eg coughing in the sense of \textit{`hello, I am still here'}   \citep{Trouvain12-CNV}, or yawning/snoring in the sense of \textit{`oh, how boring'}. 
Basically, this holds for all other NVVs.
%
Especially vegetative sounds  but also affective sounds can  be  universal. \textit{Ouch} as  expression of one's own physical pain seems to display a universal phonetic tendency in the diphthong from low/open to high/closed\footnote{``yowch'' – WordSense Online Dictionary (1st July, 2025), URL: https://www.wordsense.eu/yowch/}.
Such expressions of pain mirror the distinct bodily expressions, Darwin and James presuppose for emotions.
There is some evidence that especially negative emotions are expressed by NVVs similarly across cultures; see \cite{Sauter10-CRO,Ponsonnet24-VSI}. This might hold across species as well: \cite{Farago14-HRO} ``revealed 
similar relationships between acoustic features and emotional valence and 
intensity ratings of human and dog vocalizations''; see as well \cite{Korcsok20-ASF}.

The NVVs given in the middle of \Cref{fig:NVV-types} typically structure the dialogue and are named accordingly \emph{conversational grunts} \citep{Ward00-TCO,Ward06-NCS} or \emph{back-channeling} \citep{Yngve70-OGA,Ward06-NCS}. They often constitute rapport, \ie a good relationship, between the dialogue partners -- when they are missing or used inappropriately, this can cause irritation or outright anger. 
\emph{Filled pauses} \citep{Batliner94-API,Batliner95-FPI,Tian15-REI}    indicate hesitations but planning as well; they implicitly structure speech because mostly, they are found at specific places (not word- or phrase internally). Moreover, they characterise speakers \citep{Braun20-NVA} but can, the same way as back-channel signals, evoke affective states as well; for instance, too many filled pauses can distract the hearer and evoke irritation or even anger. 

\emph{Secondary interjections} are words and  function the same way as  primary interjections.
 This can be words/phrases with some specific affective meaning such as \emph{alas} or swear words, or longer sequences such as \emph{oh, what a joy}. 
Note that such expressions can be `discrepant', \ie not employed in their literal meaning
but indicating sarcasm/irony. 

\emph{Interjections} as word-like entities have 
been dealt with 
within linguistics and pragmatics; both primary and secondary interjections -- and \emph{semi-verbals} as fuzzy category in between -- can express  a mental attitude or state \citep{Ameka92-ITU,Wierzbicka92-TSO}, functioning as complete speech acts \citep{Poggi09-TLO}.
`Expressive interjections' are conventionalised lexical forms \citep{Ponsonnet25-TST}.
In general, the 
form of NVVs expressing affective states or traits is more or less similar to the one of verbal expressions: For instance, prototypical for anger in both is harsh voice, higher volume, higher pitch; prototypical for sadness in both is low volume, low pitch, and creaky voice.  
\cite{Lausen20-ERA} report, based on perception experiments, that listeners could classify affect bursts more accurately and with more confidence than 
speech-embedded stimuli. 
Similar results can be found in \cite{Hawk09-WAT}: 
`` ... affect vocalizations showed superior decoding over the speech stimuli for anger, contempt, disgust, fear, joy, and sadness.''
 \cite{Holz21-TPR,Holz22-TVI} showed  that a too pronounced form of affect bursts (\ie exaggerated prosody) yields lower human recognition rates.
The caveat has to be made, that all these stimuli were prompted, \ie produced intentionally. The few studies on the differences between acted and non-acted emotions indicate less variability \citep{Batliner00-DSE} and more extreme perception \citep{Barkhuysen07-CPO} for acted spoken emotions, and especially 
differences in voice quality 
\citep{Juergens11-AAP}.


\section{Functions of NVVs}
\label{sec:functions}

In the last section, we have placed  the prototypical NVVs,  affect bursts, into the larger context of interjections and verbal equivalents.
Note that we  only deal with those phenomena that can be modelled the same way as words, \ie [+vocal,-verbal], and not with those that are only concomitant with or modulated onto words, \ie [+vocal,+verbal], be this sole intonation (pitch contour), other prosodic means such as duration or intensity,  or phonation types / voice quality (modal voice, creak, harshness, breathiness, etc.), see   \cite{Laver80-TPD,Kreiman11-FOV}. Yet, such \emph{supra-segmental} features   can indicate and modify the  meaning and functions of  NVVs as well: 
NVVs are not indivisible entities but can be analysed as `having prosody' and as having segmental structure, the same way as words have. 
NVVs are multifunctional; the same segmental form (sequence of phones) can indicate different affective (or other) functions -- and of course, vice versa. Behavioural contexts can be inferred from NVVs \citep{Kamiloglu24-SLA}.
Additionally, NVVs are not only a means for expressing affects but -- even if not intended -- can evoke affective reaction on part of the receiver who is, in turn,   reacting more or less emotionally. 
Forms and functions of NVVs are interrelated:  
``Human non-verbal vocalisations thus largely parallel the
form-function mapping found in the affective calls of other animals ...'' \citep{Pisanski22-FFF}.
Yet, as far as call type is concerned, the relationship  is  ``systematic but non-redundant: listeners associated every call type with a limited, but in some cases relatively wide, range of emotions.'' \citep{Anikin18-HNV}.

The bulk of research on NVVs 
concentrates 
on their affective functions 
 -- but almost exclusively on the production of affects and their perceptual evaluation. 
 Yet, affectivism \citep{Dukes21-TRO} 
 in a full understanding had to model perception as well. Thus, we reasonably cannot restrict ourselves to emotional cries and bursts that denote pure connotations but we have to take into account all NVVs: First of all, because in real-life data, we do not encounter them with an `affective tag' but have to find out whether they are affective or not; secondly, only when taking into account 
a full interaction, we can find out whether and how an NVV evokes affect or not.

\begin{figure*}[tb!]
	\begin{center}
 \vspace{-3cm}
		\includegraphics[width=1.0\textwidth]{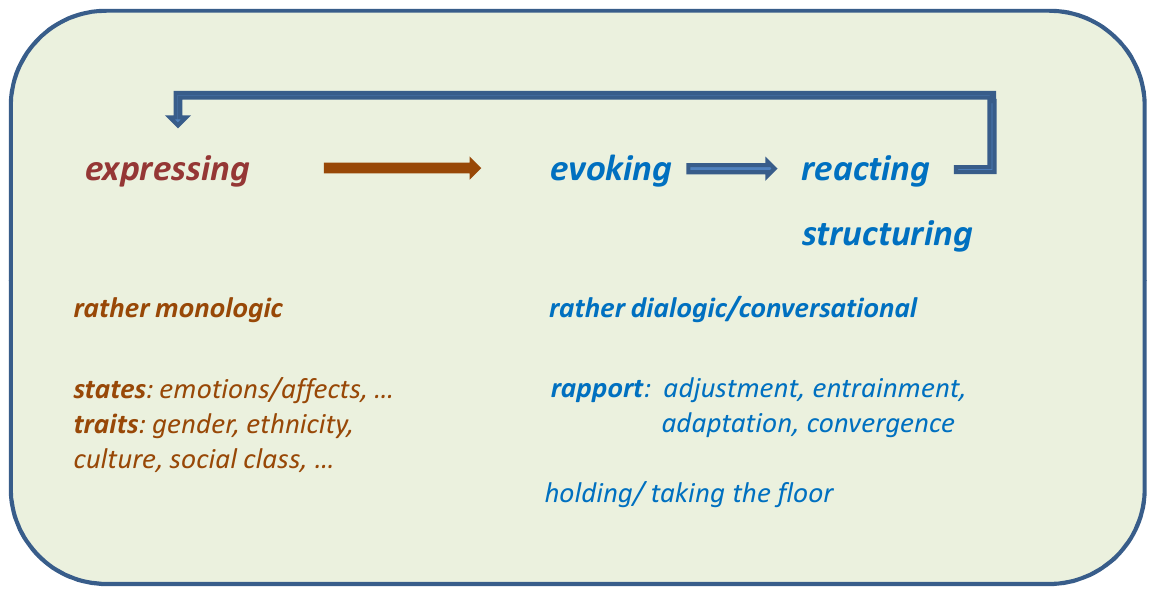}
\vspace{-3cm}
		\caption{
  Communicative functions of NVVs, ± voluntary 
  }
		\label{fig:NVV-functions}
	\end{center}
\end{figure*}

\Cref{fig:NVV-functions}  displays the main types of functions of NVVs.
Prototypically, with NVVs we express our affects but less voluntarily, we express as well (denote implicitly) all traits that can be ascribed to us such as gender, ethnicity, culture, social class. Both voluntarily as well as involuntarily expressed states and traits  evoke some reaction on the part of the interlocutor, be it affective or not. Moreover, in a dialogue, NVVs often structure in a reciprocal way. 
They help establish rapport -- there are different terms in use: adjustment, entrainment, (mutual) adaptation, and convergence.  The same way as 
 \textbf{phatic} communication 
 which rather serves a social and not a semantic function,  
rapport  is not unidirectional but reciprocal, establishing itself and changing during the course of the interaction.

\begin{figure*}[tb!]
	\begin{center}
 \vspace{-2cm}
		\includegraphics[width=1.0\textwidth]{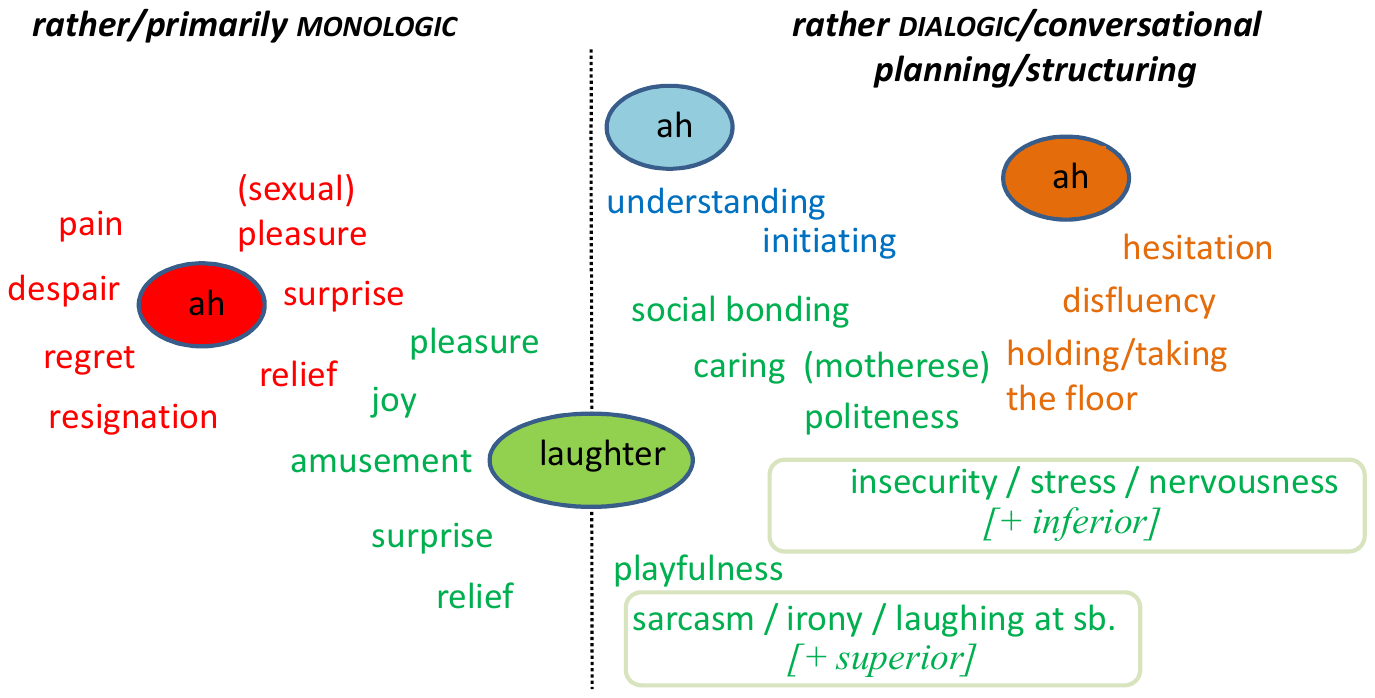}
\vspace{-4cm}
		\caption{
   The many facets of NVVs -- \textit{ah} and laughter: same or similar segmental shape, different functions expressed prosodically
  }
		\label{fig:NVV-ah}
	\end{center}
\end{figure*}

A full account of all functions of NVVs is beyond the scope of this contribution, nor can we detail the cross-cultural aspects \citep{Gendron14-CRI,Bryant21-VCA}.
We   will concentrate on three prototypical types of \textit{ah} depicted in 
\Cref{fig:NVV-ah} and illustrate  the diverse  roles of NVVs in the communication process: 

\begin{itemize}
    \item the \emph{primarily  monologic} function of expressing emotional states with \emph{affective sounds}
    \item the \emph{communicative} functions of \emph{laughter} -- a special type of affective sounds 
    \item the \emph{structuring function} of \emph{filled pauses} plus their (possible) role in evoking  reactions in the communication partner, for instance by signalling speaker idiosyncrasies
\end{itemize}


All functions depicted in \Cref{fig:NVV-ah} can  be expressed in  HHI. Those 
 left of the dotted line can be as well expressed when alone -- we do not need an interlocutor; they are primarily monologic. Those right of the dotted line are normally employed in an HHI -- they are rather dialogic. 
The phonetic form of these three NVVs needs  not but can be identical/very similar: in the case of \textit{ah}, an open long vowel \emph{[a:]}; in the case of laughter, typically with repetitive bouts \emph{hahaha} but as well with only one \emph{ha} \citep{Carus98-OTP}. 
Typical laughter is, of course, distinct from the NVV \emph{ah}; yet, in some cases, only the pragmatic context might disambiguate. 
It is not trivial to tell the three main functions and all their sub-types depicted in \Cref{fig:NVV-ah} apart, especially  in the case of real-life data, both for phonetically identical or similar forms and for different forms such as \emph{ouh, oh, ehm, uhm, mm}. Note that here, we are agnostic as for a `definite' set of functions and terms -- cover terms, sub-classes, or (near-) synonyms; such a taxonomy has 
been central for the scientific discourse 
but might unduly narrow down the possibilities when faced with our `open world scenario'.
\Cref{fig:NVV-ah} and the following three short descriptions provide just an  excerpt out of the plethora of different forms and functions of NVVs; thus, they are not intended as full-fledged taxonomies. 





\paragraph{Affective sounds:}

Within the paradigm of the `big n' emotions, \cite{Simon09-TVC} extended the traditionally limited set  (\emph{affect bursts} in Scherer's conceptualisation) ``anger, disgust, fear, sadness, surprise, happiness, and for the voice, also tenderness.''
They claim that brief vocal bursts can communicate 22 different 
emotions -- nine negative and thirteen positive. At the same time, they stress that this is not a ``definite nor exhaustive set of emotions with specific vocalizations''.
%
\cite{Cowen19-M2E} assume that ``at least 24 distinct kinds of emotion [...] conveyed by vocal bursts are bridged by smooth gradients with continuously varying meaning.'' 
This extension of the basic emotions and its theoretic conceptualisation were criticised by \cite{Crivelli19-IFB};  see the reply by \cite{Keltner19-WBE}.
%
Basically, it is still not settled which of such  categories are really distinct and to which extent there are more -- and more or less distinct -- categories.
In \Cref{fig:NVV-ah}, we display in red some important affective functions  that can be indicated by \emph{ah}, coarsely arranged along the arousal dimension (high above, low below) and the valence dimension (negative left, positive right); this is not exhaustive -- we can debate which of the 213 `emotion words', \ie emotion (sub-)categories,  given by   \cite{Shaver87-EKF}, can as well be expressed the same way. It might be difficult to tell apart a purely communicative function of \emph{understanding} (given in blue) from more affective connotations such as \emph{surprise}, and from the use as conversational strategy (intitiating interaction) -- or even from \emph{ah} as pure hesitation (given in orange).
A (linguistic/situational) context can disambiguate; in the case of stand-alone NVVs, this might be less  possible, though.




\paragraph{Laughter:}

In \Cref{fig:NVV-ah}, the functions of laughter (in green) are coarsely ordered by being primarily monologic (left) and being primarily dialogic/interactive (right). 
A  general overview is given by \cite{Trouvain17-L}; as for the different forms of laughter, see  \cite{Bacharowski01-NAL,Bacharowski01-TAF}. Of note is its  special  role in the parent/child interaction (motherese/parentese, see \cite{Fernald94-HMV}.
Laughter mirrors  syntax \citep{Provine93-LPS,Batliner19-OLA} and conversation/discourse \citep{Bonin14-TFL,Gavioli95-TVT,Ludusan22-LEI,Mazzocconi22-WYL,Vettin04-LIC}. 
Laughter constructs meaning, identities, and relationships in social interaction \citep{Rees10-ISB}.
\cite{Rychlowska22-TRO} stress the role of context for classifying the functions of laughter.
Note that we can employ laughter from a superior or from an inferior stance; perceptual confusions, especially those pertaining to these contrasting stances  -- made by humans or by machines alike -- can be very critical. `Voluntary confusions' -- when a subordinate is sarcastic, by that assuming a superior stance -- will evoke specific effects as well.


\paragraph{Filled pauses:}
Hesitations (unfilled or filled pauses, etc., marked orange in  \Cref{fig:NVV-functions}) are, from the point of view of `correct' grammar, un- or dys-grammatical phenomena (disfluencies); a full account of such disfluencies is given in 
\cite{Batliner94-API}. 
Filled pauses \citep{Batliner95-FPI} -- which normally are conceived as `non-affective' NVVs  that characterise speaker idiosyncrasies and planning strategies -- can influence the perception of personality: 
A lower amount of disfluencies may make a speaker appear more confident and focused \citep{Kirkland23-PMD}. In general, disfluencies can be harnessed for forensic purposes \citep{Braun20-NVA} -- frequencies and segmental-prosodic shape can, for instance, characterise and verify speakers; 
moreover, they can provide additional informations for recognising emotions in dialogues \citep{Tian15-REI}. Typically, they are found in interaction but are rather `speaker-oriented than hearer-oriented' and not `prone to conversational convergence' \citep{Hutin24-UUA}.

\paragraph{}
The bulk of studies on NVVs has been conducted for HHI. 
So far, studies on HRI are only a few, concentrating on specific NVVs such as laughter; see Sec. \ref{sec:computational-approaches}.
Note that the tendency to deal with NVVs out of context -- as if they had been produced in isolation -- does not mirror real life: ``They are responsive to prior utterances or elicit responses in turn'' \citep{Dingemanse17-OTM}.

\vspace{2cm}

\section{Basic approaches: Data, methods, and frequencies}
\label{sec:basic-approaches}

\begin{figure*}[tb!]
	\begin{center}
 \vspace{-2cm}
		\includegraphics[width=1.0\textwidth]{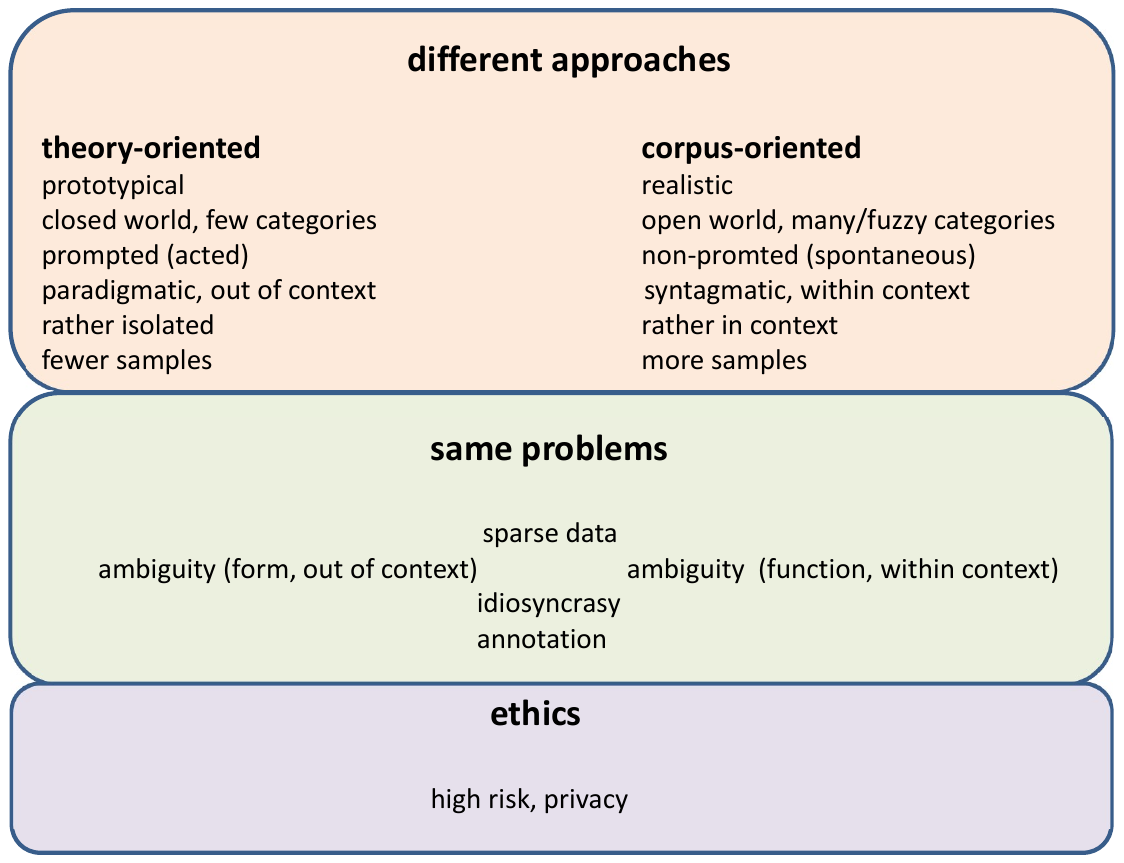}
\vspace{-2cm}
		\caption{
  Approaches: data, methods, and problems
  }
		\label{fig:NVV-approaches}
	\end{center}
\end{figure*}

In this section, we will contrast the two different approaches towards data and their collection in speech and language and by that, in studies on NVVs as well: \emph{theory-oriented} vs \emph{corpus-oriented} approaches. A summary is given in \Cref{fig:NVV-approaches}. 
As already mentioned in  \Cref{sec:introduction}, we use `databases' for collections of (pre-defined/selected) items, and `corpora' for collections of not pre-defined/selected items within a (real-life) context.
Typical and by that, formative for NVVs are the 
\emph{theory-oriented} data collections  starting in the 90ies inspired by  (basic) emotion theories. Real-life data served as inspiration, but the stimuli were carefully defined and prompted in the lab; due to this effort, the number of items per class was rather low. 


Another line of research was motivated by the need for larger databases for the training of ASR and then, for speech in context (\eg dialogues or multi-party conversations),  
see \cite{Trouvain12-CNV}. 
This typically resulted in  corpora with durations spanning from some hours to several hundred hours. 
NVVs were rather a by-product, differed considerably in frequencies, and were not always systematically taken into account.
In the last years, hybrid approaches came to the fore, \eg prompted data collected `in the wild',  over the web and possibly embedded into a context. Yet, they still lack the characteristics listed for corpus-oriented approaches in \Cref{fig:NVV-approaches}.
%
A consistent shortcoming of both theory- and corpus-oriented approaches is the low number of NVVs, especially given the need for scaling in modern AI approaches.
To illustrate this shortcoming,  we will now report characteristic sizes of databases targeting NVVs, as well as the number of NVVs found in corpora that were not especially aimed at collecting NVVs.

Emotional NVV tokens in studies can comprise as few as 36 \citep{Gendron14-CRI} and up to a few hundred \citep{Anikin17-PAA}. 
Non-prompted, `spontaneous’ NVVs obtained from specific scenarios are normally on a similar scale, \eg 968 hesitations (filled pauses) in \cite{Batliner95-FPI} or 176 laughters in \cite{Batliner19-OLA}. 
 Only in a few large-scale data collections, there are markedly more tokens: Of note are  the over 1500 hours of conversations in a private setting  \citep{Campbell02-RTF,Campbell04-SET}  which contain more than 10\% non-verbals/laughters, and the few large-scale collections of multiparty meetings: laughs $>12k$  in the ICSI corpus (72 hours), $>12k$  in the AMI corpus (100 hours), and $>22k$  in the Switchboard corpus (518 hours), see \cite{Trouvain12-CNV}. Laughter seems to be the only NVV that could be  found in a sufficient order of magnitude. As deep learning approaches cannot reasonably be trained with only a few items per class, \citet{Keltner19-EEA,Keltner19-WBE} 
 resorted to convenience sampling obtained via the web, resulting in  $>60k$ NVVs. 
A similar dataset is EmoGator, consisting of some $32k$ items, 30 distinct categories, and 357 speakers, obtained from volunteers and crowd-sourced workers, see \cite{Buhl23-EAN}; these data have been employed by \cite{Maharjan24-STI}. 
\cite{Norrick14-I} lists the `Most frequent initial and free-standing
interjections in LSWE-AC'  (the Longman Spoken and Written English corpus, AE conversation corpus) with 329 texts and 2,480,800 words): Most frequent is \textit{yeah} with 40,652 tokens, followed by \textit{oh} with 28,380 tokens, \textit{um, uh} together 3,803+3,608 $=$ 7,411 tokens, and \textit{ah} 846 tokens; note that these are written sources.  
Dingemanse reports for 1.3 million turns of speech in 18 languages that interjections occur every 12\,s, constituting one out of every seven turns \citep{Dingemanse24-IAT}, but that -- in a corpus of spoken Dutch -- only about 7\% ``was expressive of the speaker's mental or affective state.'' \citep{Dingemanse23-I}. 
This distribution is in line with \cite{Goddard14-IAE}  who pointed out that in large corpora, affective NVVs are underrepresented while discursive NVVs are overrepresented.
This very fact illustrates the dilemma: Collecting and annotating such large databases is costly and does not necessarily result in enough items per NVV class. Thus, researchers tend to elicit (prompt for) NVVs -- or, in the case of data collected over the web (YouTube, movies), to preselect.
This means, however, that all these items are most of the time acted/staged -- we do not know up to what extent we can find them in real life. Not only are such prompted data out of context,  \cite{Anikin17-PAA} showed that listeners can distinguish between authentic and acted emotions -- 
the form of NVVs we encounter in real-life data is, somehow and sometimes, different from the one of prompted data. 
(On a side note: Strictly speaking, we only can talk of different speaking styles (registers), not of, \eg `acted' vs `spontaneous'. Yet, it should be clear that `acted'/`prompted' data are not representative for `spontaneous' speech in real-life scenarios.)

The main and the same problems to be addressed that hold for both theory-oriented and data-oriented approaches alike are given  in 
\Cref{fig:NVV-approaches}, below. As far as methodology is concerned, we not only have to face very sparse data when we aim at realistic data; speaker-idiosyncratic uses and, partly due to that, ambiguities can only be solved when we can model a large enough context. And this has not yet been done. Moreover, especially strongly affective NVVs might be found more often in non-transactional, rather private contexts. This poses severe problems due to ethical  concerns -- depending on the application envisioned \citep{Batliner23-EAI} -- and explains why such data collections such as the one of \cite{Campbell02-RTF,Campbell02-TAG,Campbell04-SET} are rather singular. 
Apart from private encounters, strong emotions -- and by that, NVVs expressing strong emotions -- might be found in critical situations in public spaces; (automatic) surveillance of such places seems to be meaningful but might be in conflict with ethics regulations, \eg in the EU AI act \citep{EU2021-AI}.


Existence determines consciousness,  and doing determines theories: The presumed necessity and possibility of collecting isolated NVVs at large scale favours the concept of NVVs as primary interjections, produced in isolation. In part, this holds for speech emotion recognition (SER) as well; yet,  there continuous annotation of emotion dimensions makes it at least possible to annotate and model not only snippets but longer stretches of speech and by that, conversations, as well.
On the one hand, selected prompted NVVs  
 do not necessarily 
 mirror real-life data but are only claimed to model them. 
 On the other hand, specific (real-life) scenarios only provide a small selection out of all NVV  categories and classes. 
 We might be able to record and process data obtained from interactions between conversation partners that are, however, only rarely  full-blown emotional. Moreover, models trained within (too few) very private settings might not be allowed to be used at all because there will be serious ethical concerns; data are too sparse, and emotions too private. Thus, we might be forced to target more `transactional' (public), \ie less private settings which are non-prompted (non-acted) but sort of `realistic'. 
\cite{Shiota23-TFO} rightly mention  the huge gap  ``between affect/emotion in the lab and in real life'' and argue  against easy-to-use (convenience) samples obtained via the web:   ``In moving beyond the lab, we also caution against over-relying on cheap, easy-to-recruit online samples and questionnaire measures (e.g., via 
MTurk and Prolific)'', 
because it ``is arguably even more impoverished than laboratory research at 
capturing real-life experience''; 
as for other studies critical towards (cross-cultural) crowd sourced data, see \cite{Fort14-CFL,Gvirtz24-TLO}.
Moreover, as often, research so far has concentrated on the `WEIRD' societies \citep{Henrich10-TWP} 
and has only rarely taken into account cross-cultural aspects, see \cite{Shiota23-TFO}.

\section{Computational approaches towards NVVs}
\label{sec:computational-approaches}


Classification performance depends on (type of) data, size of database, and (type of) computational procedure used. 
Most of the time, for NVVs, classification has been so far  out-of-context, \ie isolated NVVs are processed. We do not know of any study where each (all types of) NVVs have been processed within context (\ie in an open world scenario) in a detection and classification paradigm. This has been done only for specific NVVs such as laughter, see \cite{Batliner19-OLA}, or for a restricted set of NVVs, \eg laughter and shouts, together with six `big' emotions, in \cite{Hsu21-SER}. 
%
Overviews of NVVs for Human-Robot-Interaction (HRI) are given in \cite{Yilmazyildiz16-ROS} (called `semantic free utterances') and \cite{Zhang23-NSI}. 
\cite{Haddad16-LAS} concentrate on laughter and smile in HRI  
and discuss the benefit of including them for 
improving the quality of the interaction and helping in modelling the user's state.

A detection task is addressed in \cite{Tzirakis23-LNV}.
The authors obtain impressive UAR values of more than 90\% correct. 
They do not, however, employ data from a realistic scenario but overlay  NVVs obtained from crowd-sourcing with noisification and speech from other databases, within a balanced design. By that, they avoid many of the problems we pointed out. Moreover,  such a combination from diverse sources can be prone to `clever Hans' problems, \ie detection and classification run risk to be based on wrong proxies (confounders);
cf., for instance,  \cite{Coppock24-AAC-long} 
who  demonstrated for another task (detection of Covid 19 in crowd-sourced data) a pronounced  degradation  when  confounders are taken into account.



Measures used for automatic classification have been so far accuracy and, for unbalanced classes, mostly unweighted average recall (UAR) or correlations. The 44\% UAR of the winner of the 
`vocalisation challenge' at ACM-MM 2022 
\citep{Schuller22-TAM}, obtained with Wav2Vec2 \citep{Grosz22-WPS} for six classes, can give an impression of present-day performance -- with due caveats, because of the small number of items and the specificities of the data. 
\cite{Koudounas25-VAF} introduce a new foundation model, voc2vec, claiming better performance for available open source data sets with vocalisations.
Although NVVs can be stand-alone and had to be modelled and recognised in such cases, it is more meaningful to model them in their (linguistic) context.
 NVVs do normally not occur in isolation 
 but in the context of  speech -- especially in scenarios where we can expect that recordings of such real-life data can be employed because they do not violate privacy restrictions.

Now, we sketch 
alternatives to collecting pre-selected, prompted, out-of-context items. 
First, we can go the slow path of corpus-based research, \ie address non-prompted speech data in specific scenarios, not necessarily aimed at collecting NVVs; they will be a by-product and mostly sparse, though.
Yet, some of these scenarios might really provide  higher numbers of NVVs, especially affect bursts, \eg  video games.
%
Second, we can concentrate on generation/synthesis of specific NVVs within specific scenarios,  \eg laughter to relax communication, or hesitations as indicators of a system that contemplates and by that gives a more `human' impression. 
A coarser modelling  of the well-known dimensions valence and arousal might help avoiding misses and false alarms. Again, this might be meaningful for specific scenarios.
%

Third, within Artificial Intelligence (AI), 
we  can address  NVVs with state-of-the-art foundation models. 
This process includes a multi-stage pipeline integrating audio processing, contextual reasoning, and 
understanding. In the initial stage, NVVs are segmented from the audio stream using foundation models such as Whisper~\citep{Radford22-RSR} or WavLM~\citep{Chen21-WLS}.
Once NVVs are isolated,  latent audio embeddings using self-supervised models are extracted such as Wav2Vec 2.0~\citep{Baevski20-WAF}, HuBERT~\citep{Hsu21-HSS}, or ExHuBERT~\citep{amiriparian2024exhubert}. These models generate high-dimensional representations that capture rich acoustic and paralinguistic features, potentially encoding prosody including voice quality and speaker affect.
These embeddings can then be used to classify NVVs into specific categories (\eg sighs, laughs, gasps) and mapped to emotional states using models based, \eg on arousal-valence dimensions. 
Now, by integrating NVV embeddings, their inferred emotional tags, and the surrounding linguistic context, LLMs can better interpret the communicative function of NVVs. 
This contextual reasoning is essential for disambiguating NVVs, whose meaning often depends heavily on the speaker's intent, prior dialogue, and social dynamics.

Let's now use our quotation from Hamlet  \textit{``Ah, Rosencrantz! Good lads, how do ye both?''} as example: We can imagine that the LLM receives the textual utterance, any relevant preceding dialogue (e.g., Hamlet has not seen Rosencrantz for some time), and the NVV embedding and its associated affective interpretation. The LLM can then synthesize these inputs to produce a contextualized reading of the NVV: The speaker utters ``Ah'' with a soft, mid-pitched tone that signals warmth and familiarity. Given the subsequent context, the LLM then concludes that the NVV most likely expresses positive surprise  and fondness
upon encountering familiar friends. 
Note that this interpretation might stick to the literal meaning and disregard the sarcasm behind it.
Now  Shakespeare's work and its interpretations are surely in the training data of LLMs, so all this exemplifies  their capabilities as if  but cannot prove them.



The ability to perform in-context reasoning makes LLMs a good option for interpreting NVVs in dynamic or open-ended settings, where rigid, predefined mappings from acoustic features to meaning may fall short. Rather than relying solely on fixed taxonomies or labelled datasets, LLMs can interpret NVVs based on how they function in discourse, how they relate to past and future utterances, and how they align with known patterns of social and emotional behavior. Yet, the caveat has to be made that LLMs have still to demonstrate  this  suitability for modelling NVVs in real-life data.

\vspace{1cm}

\section{Concluding remarks}
\label{sec:concluding-remarks}

In this contribution, we presented NVVs and their various forms and functions, on an exemplary basis.
We  elaborated on the difficulties we are faced when trying to collect and model them in a realistic way, aiming at harnessing them in machine learning devices for HCI. 
The biggest problem might be that NVVs are ubiquitous as a whole but too often sparse in both theory-based and corpus-based data collections, if it comes to specific functions and forms. Thus, researchers have resorted to prompted NVVs and/or to collect them out-of-context; this facilitates the use of AI procedures but fully ignores the syntagmatic aspect -- when, where, towards whom they have been employed by whom, and even the paradigmatic aspect -- their acoustic form (phonetics, prosody) and their segmental structure in different contexts. (Note that we do not argue against constructed/selected stimuli used in basic research, when a specific hypothesis is addressed which requires a strict \textit{ceteris paribus}).

In analogy to the term \emph{Model Autophagy Disorder} (MAD) \citep{Alemohammad23-SGM} that describes the unfavourable consequences of training AI with AI generated content, we can introduce the term  \emph{Model Malnutrition Disorder} (MMD) for the unfavourable consequences of feeding models with non/less appropriate (acted or selected) data.
Training datasets 
`` shape the epistemic boundaries governing how AI operates and, in that sense, create the limits of how AI can `see' the world.'' \citep[p. 98]{Crawford21-AOA}.
They constitute a \emph{closed world} -- where AI can show off its strengths, rather different from the \emph{open world} we should model when we aim at real-life applications. 
So far, it is not possible to demonstrate the effect of MMD because we are lacking (very) large  realistic databases containing enough instances of NVVs that could constitute an upper baseline for performance measures; 
and we do know that realistic NVVs behave differently from acted NVVs -- but we do not know how much.


We can conceive NVVs as the dressing that makes dishes special -- not necessarily needed but, at the same time, pivotal. 
They are much more than just some prototypical affect bursts; they are ubiquitous, and they cannot be attributed simply to a few mutually exclusive cover classes such as [$\pm$affective]. In the literature, so far (too) much weight has been put on the production/generation aspect, focusing on either affects or conversational phenomena. Equally important is  perception,  \ie which affects are triggered by  NVVs that have not been intended to indicate affect but are simply indicating speaker idiosyncrasies,  are produced involuntarily, or are employed as `affective signals' without being affective per se. 

As  mentioned in Section \ref{sec:history-and-definitions}, NVVs have always been seen as something special. Yet, maybe we should turn the tables: Even  affective NVVs need not be seen as  special but can be treated the same way as words, albeit with a high functional and pragmatical load. They normally do not contradict their linguistic and situational context; and if they do, they do it the same way as words would do. We can analyse and interpret NVVs employing their context -- and vice versa. Only in the  specific cases when NVVs stand fully alone  -- for instance, pre-linguistic babies communicate only non-verbal, we have to analyse them by their own, the same way as we had to analyse a one-word utterance. 
NVVs  might be the preferred means of communication for fighting, fear of life, or sexual intercourse \citep{Anikin24-WDP}; yet, in all these situations, adults can and do employ speech as well. Such situations are definitely interesting objects of scientific investigations. They might lead to genuine application scenarios, excluding linguistic markers,  in the case of  sex robots   or  automatic surveillance systems.
But for such scenarios and in general, we must not ignore the `ethics of NVVs': the more affective they are, the higher is the probability that they are found within and indicate a rather private scenario. Thus, they have to be embedded into a wider context of ethical considerations on modelling and harnessing them in prospective applications.

\paragraph{Acknowledgements}
This work received funding from the German Research Foundation (DFG; Reinhart Koselleck project AUDI0NOMOUS, No. 442218748).

\newpage


\bibliography{NVV}



\end{document}